\documentclass[11pt,letterpaper]{article}

\usepackage{amsmath}
\usepackage{amssymb}
\usepackage{amsthm}
\usepackage{fullpage}
\usepackage[utf8]{inputenc}
\usepackage{algpseudocode}
\usepackage{algorithm}
\usepackage{newclude}
\usepackage{lmodern}
\usepackage[T1]{fontenc}
\usepackage{hyperref}

\newtheorem{theorem}{Theorem}
\newtheorem{lemma}[theorem]{Lemma}
\newtheorem{definition}[theorem]{Definition}
\newtheorem{corollary}[theorem]{Corollary}

\newcommand{\set}[1]{\left\{#1\right\}}
\newcommand{\Prb}[2]{\mathrm{Pr}_{#1}\left[#2\right]}

\newcommand{\Nbb}{\mathbb{N}}
\newcommand{\Rbb}{\mathbb{R}}
\newcommand{\Zbb}{\mathbb{Z}}
\newcommand{\eps}{\varepsilon}
\renewcommand{\paragraph}[1]{{\bf #1.}}
\newcommand{\Ac}{\mathcal{A}}
\newcommand{\Bc}{\mathcal{B}}
\newcommand{\Cc}{\mathcal{C}}
\newcommand{\Dc}{\mathcal{D}}
\newcommand{\Uc}{\mathcal{U}}
\newcommand{\Vc}{\mathcal{V}}

\begin{document}
    \title{Beyond Locality-Sensitive Hashing}
    \author{Alexandr Andoni\\ Microsoft Research \\ \texttt{\small andoni@microsoft.com} \and Piotr Indyk\\ MIT\\\texttt{\small indyk@mit.edu} \and Huy L. Nguy\~{\^{e}}n\\ Princeton \\ \texttt{\small hlnguyen@princeton.edu} \and Ilya Razenshteyn \\ MIT \\ \texttt{\small ilyaraz@mit.edu}}
    \date{\vspace{-5ex}}
    \maketitle
    \begin{abstract}
        We present a new data structure for the $c$-approximate near neighbor problem (ANN) in the Euclidean space.
        For $n$ points in $\Rbb^d$, our algorithm achieves $O_c(n^{\rho} + d \log n)$ query time and $O_c(n^{1 + \rho} + d \log n)$ space, where $\rho \leq 7/(8c^2) + O(1 / c^3) + o_c(1)$. 
        This is the first improvement over the result by Andoni and Indyk (FOCS~2006) and
        the first data structure that bypasses a locality-sensitive hashing lower bound
        proved by O'Donnell, Wu and Zhou (ICS~2011).
        By a standard reduction we obtain a data structure for the Hamming space and $\ell_1$ norm with
        $\rho \leq 7/(8c) + O(1/c^{3/2}) + o_c(1)$, which is the first improvement over the
        result of Indyk and Motwani (STOC~1998).
    \end{abstract}


    \section{Introduction}

The near neighbor search problem  is defined as follows:
given a set $P$ of $n$ points in a $d$-dimensional space,  build a data structure that, given a query point $q$,  reports any point within a given distance $r$
to the query (if one exists). 
The problem is of major importance in several
areas, such as databases
and data mining, information retrieval, computer vision, 
databases and signal processing.

Many efficient near(est) neighbor  algorithms are known for the case when the dimension $d$ is ``low'' (e.g., see~\cite{m-plah-93}, building on~\cite{c-racpq-88}). 
However, despite decades of effort, the
current solutions suffer from either space or query time that
are exponential in the dimension $d$. This phenomenon is often called ``the curse of dimensionality''.
 To overcome this state of affairs, several researchers proposed {\em approximation} algorithms for the problem.
In the $(c, r)$-approximate near neighbor problem (ANN), the data structure is allowed to
return any data point whose distance from the query is at most $c r$, for an approximation factor $c > 1$.  Many approximation algorithms for the problem are known, offering tradeoffs between the approximation factor, the space and the query time. See~\cite{a-nnson-09} for an up to date survey. 

From the practical perspective, the space used by an algorithm should be as close to linear as possible. If the space bound is (say) sub-quadratic,  and the approximation factor $c$ is a constant, the best existing solutions are based on {\em locality-sensitive hashing}~\cite{im-anntr-98,him-anntr-12}. 
The idea of that approach is to hash the points in a way that the probability of collision is much higher for points which are close (with the distance $r$) to each other than for those which are far apart (with distance at least $cr$). Given such hash functions, one can retrieve near neighbors by hashing the query point and retrieving elements stored in buckets containing that point. 
  If the probability of collision is at least $p_1$ for the close points and at most $p_2$ for the far points, the algorithm solves the $(c, r)$-ANN using $n^{1+\rho+o(1)}$ extra space and $dn^{\rho+o(1)}$ query time\footnote{Assuming that each hash function can be sampled and evaluated in $n^{o(1)}$ time, stored in $n^{o(1)}$ space, that distances can be computed in $O(d)$ time, and that $1/p_1 =n^{o(1)}$.}, where $\rho=\log(1/p_1)/\log(1/p_2)$~\cite{him-anntr-12}. The value of the exponent $\rho$ depends on the distance function and the locality-sensitive hash functions used.
 In particular, it is possible to achieve $\rho=1/c$ for the $\ell_1$ norm~\cite{im-anntr-98}, and $\rho=1/c^2+o_c(1)$ for the 
 $\ell_2$ norm~\cite{ai-nohaa-06}.
 
It is known that   the above bounds for the value of $\rho$ are {\em  tight.} Specifically, we have that, for all values of $c$, $\rho \ge 1/c - o_c(1)$ for the $\ell_1$ norm\footnote{Assuming $1/p_1 =n^{o(1)}$.}~\cite{owz-olbls-11}. A straightforward reduction implies that $\rho \ge 1/c^2 - o_c(1)$ for the $\ell_2$ norm. Thus, the running time of the simple LSH-based algorithm, which is determined by $\rho$, cannot be improved.

\paragraph{Results} In this paper we show that, despite the aforementioned limitation,  the space and query time bounds for ANN can be substantially improved. In particular, for the $\ell_2$ norm, we give an algorithm with query time $O_c(n^{\eta} + d \log n)$ and space $O_c(n^{1+\eta} + d \log n)$, where $\eta =\eta(c) \le 7/(8 c^2) + O(1/c^3) + o_c(1)$ that gives an improvement for large enough $c$. This also implies an algorithm with the exponent $\eta \le 7/(8 c) + O(1/c^{3/2}) + o_c(1)$ for the $\ell_1$ norm, by a classic reduction from $\ell_1$ to $\ell_2$-squared~\cite{llr-ggsaa-95}. These results constitute the first improvement to the complexity of the problem since the works of \cite{im-anntr-98} and~\cite{ai-nohaa-06}.

\paragraph{Techniques} Perhaps surprisingly, our results are obtained by using essentially the same LSH functions families as described in~\cite{ai-nohaa-06} or~\cite{im-anntr-98}. However, the properties of those hash functions that we exploit, as well as the overall algorithm, are different. On a high-level, our algorithms are obtained by combining the following two observations:
\begin{enumerate}
\item After a slight modification, the existing LSH functions can
  yield better values of the exponent $\rho$ if the search radius $r$
  is comparable to the diameter\footnote{In the analysis we use a
    notion that is weaker than the diameter. However, we ignore this
    detail for now for the sake of clarity.} of the point-set. This is
  achieved by augmenting those functions with a ``center point''
  around which the hashing is performed.
  See Section~\ref{ss:int} for an intuition why this approach works, in the (somewhat simpler) context of the Hamming distance.
\item We can ensure that the diameter of the point-set is small by
  applying standard LSH functions to the original point-set $P$, and
  building a separate data structure for each bucket.
\end{enumerate}

This approach leads to a two-level hashing algorithm. The {\em outer
  hash table} partitions the data sets into buckets of bounded
diameter. Then, for each bucket, we build the {\em inner hash table},
which uses (after some pruning) the center of the minimum enclosing
ball of the points in the bucket as a center point.  Note that the
resulting two-level hash functions cannot be ``unwrapped'' to yield a
standard LSH family, as each bucket uses slightly different LSH
functions, parametrized by different center points. That is, the
two-level hashing is done in a {\em data-aware} manner
while the
standard LSH functions are chosen from a distribution independent from
the data. This enables us to overcome the lower bound of~\cite{owz-olbls-11}.

Many or most of the practical applications of LSH involve designing data-aware hash functions. 
Unfortunately, not many rigorous results in this area are known. 
The challenge of understanding and exploiting the relative strengths of data-oblivious versus data-aware methods has been recognized as a major open question in the area (e.g., see~\cite{fmda-13}, page 77). 
Our results can be viewed as a step towards that goal.

\paragraph{Related work}
In this paper we assume worst case input.  If the input is generated
at random, it is known that one can achieve better running
times. Specifically, assume that all points are generated uniformly at
random from $\set{0,1}^d$, and the query point is ``planted'' at
distance $d/(2c)$ from its near neighbor. In this setting, the work of
\cite{cr-fflas-93, gpy-mihir-94, kwz-bvipp-95, prr-lbp-95} gives an exponent of roughly $\tfrac{1}{\ln
  4\cdot c}\approx \tfrac{1}{1.39c}$.

Even better results are known for the problem of finding the {\em
  closest pair} of points in a dataset. In particular, the algorithm
of \cite{d-bcits-10} for the random closest pair has an exponent of
$1+\tfrac{1}{2c-1}$.\footnote{Note that a near neighbor search
  algorithm with query time $n^\rho$ and space/preprocessing time of
  $n^{1+\rho}$ naturally leads to a solution for the closest pair
  problem with the runtime of $n^{1+\rho}$.} More recently, \cite{v-fcsta-12}
showed how to obtain an algorithm with a runtime exponent $<1.79$ for
any approximation $c=1+\eps$ in the random case. Moreover, \cite{v-fcsta-12}
also gives an algorithm for the worst-case closest pair problem with a
runtime exponent of $2-\Omega(\sqrt{\eps})$ for $c=1+\eps$
approximation.

There are also two related lines of lower bounds for ANN. First, the work of \cite{mnp-lblsh-07} showed that LSH for Hamming space must have $\rho\ge 1/(2c)-O(1/c^2)-o_c(1)$, and \cite{owz-olbls-11} improved the lower bound to $\rho\ge 1/c-o_c(1)$. Second, \cite{ptw-galba-08,ptw-lbnns-10} have given cell-probe lower bounds for $\ell_1$ and $\ell_2$, roughly showing that any randomized ANN algorithm for the $\ell_1$ norm must either use space $n^{1+\Omega(1/(tc))}$ or more than $t$ cell-probes. We note that the LSH lower bound of $\rho\ge 1/(2c)$ from \cite{mnp-lblsh-07} might more naturally predict lower bounds for ANN because it induces a ``hard distribution'' that corresponds to the aforementioned ``random case'' . In contrast, if one tries to generalize the LSH lower bound of \cite{owz-olbls-11} into a near neighbor hard distribution, one obtains a dataset with special structure, which one can exploit (and our algorithm will indeed exploit such structure). In fact, the LSH lower bound of \cite{mnp-lblsh-07} has been used (at least implicitly) in the data structure lower bounds from \cite{ptw-galba-08, ptw-lbnns-10}.

\subsection{Intuition behind the improvement}
\label{ss:int}
We give a brief intuition on why near neighbor instances with bounded diameter are amenable to more efficient LSH functions. 
For simplicity we consider the Hamming distance  as opposed to the Euclidean distance.  

Assume that all input points, as well as the query point, are within the Hamming distance of $s$ from each other. 
By shifting one of the data points to the origin, we can assume that all points have at most $s$ non-zeros (i.e., ones). Consider any data point $p$ and the query point $q$. To make calculations easier, we assume that both $p$ and $q$ have exactly $s$ ones.

The ``standard'' LSH functions for the Hamming distance project the points on one of the coordinates selected uniformly at random. 
For two points $p$ and $q$ this results in a collision probability of $1-\|p-q\|_1/d$, which is $1-r/d$ and $1-cr/d$ for points within the distance of $r$ and $cr$, respectively. The probability gap of $1-x$ vs. $1-cx$ leads to the exponent $\rho$ equal to $1/c$ \cite{im-anntr-98}. To improve on this, we can instead use the min-wise hash functions of~\cite{b-fndd-98}. For those functions, the probability of collision between two points $p$ and $q$ is equal to $\frac{|p\cap q|}{|p \cup q|} $, where $\cup$ and $\cap$ denote the union and intersection of two Boolean vectors, respectively. 
Since we assumed that $\|p\|_1=\|q\|_1=s$, we have

$$\frac{|p\cap q|}{|p \cup q|} = \frac{\|p\|_1 + \|q\|_1 - \|p-q\|_1}{\|p\|_1+\|q\|_1 + \|p-q\|_1} = \frac{2s- \|p-q\|_1}{2s+\|p-q\|_1}
= \frac{1-\|p-q\|_1/(2s)}{1+ \|p-q\|_1/(2s)} $$

As a result, the collision probability gap for distances $r$ and $cr$ becomes $\frac{1-x}{1+x}$ vs. $\frac{1-cx}{1+cx}$. This leads to $\rho$ that is lower than $1/c$.

    \section{Preliminaries}

In the text we denote the $\ell_2$ norm by $\| \cdot \|$. When we use $O(\cdot)$, $o(\cdot)$,
$\Omega(\cdot)$ or $\omega(\cdot)$ we explicitly write all the parameters that the corresponding constant factors
depend on as subscripts.

\begin{definition}
    The \emph{$(c, r)$-approximate near neighbor problem (ANN)} with failure
    probability $f$ is to construct a data structure over a set
    of points $P$ in metric space $(X, D)$ supporting the following
    query: given any fixed query point $q \in X$, if there exists
    $p \in P$ with $D(p, q) \leq r$, then report some $p' \in P$
    such that $D(p', q) \leq cr$, with probability at least $1 - f$.
\end{definition}
\textbf{Remark:} note that we allow preprocessing to be randomized as well,
and we measure the probability of success over the random coins tossed
during \emph{both} preprocessing and query phases.

\begin{definition}[\cite{him-anntr-12}]
    For a metric space $(X, D)$ we call a family of hash functions $\mathcal{H}$ on $X$ \emph{$(r_1, r_2, p_1, p_2)$-sensitive}, if
    for every $x, y \in X$ we have
    \begin{itemize}
        \item if $D(x, y) \leq r_1$, then $\mathrm{Pr}_{h \sim \mathcal{H}}[h(x) = h(y)] \geq p_1$;
        \item if $D(x, y) \geq r_2$, then $\mathrm{Pr}_{h \sim \mathcal{H}}[h(x) = h(y)] \leq p_2$.
    \end{itemize}
\end{definition}

\textbf{Remark:} for $\mathcal{H}$ to be useful we should have $r_1 < r_2$ and $p_1 > p_2$.

\begin{definition}
    If $\mathcal{H}$ is a family of hash functions on a metric space $X$, then for any $k \in \Nbb$ we can define a family of hash function
    $\mathcal{H}^{\otimes k}$ as follows: to sample a function from $\mathcal{H}^{\otimes k}$ we sample $k$ functions $h_1, h_2, \ldots, h_k$
    from $\mathcal{H}$ independently and map $x \in X$ to $(h_1(x), h_2(x), \ldots, h_k(x))$.
\end{definition}

\begin{lemma}
    If $\mathcal{H}$ is $(r_1, r_2, p_1, p_2)$-sensitive, then $\mathcal{H}^{\otimes k}$ is $(r_1, r_2, p_1^k, p_2^k)$-sensitive.
\end{lemma}

\begin{theorem}[\cite{him-anntr-12}]
    \label{lsh_to_nn}
    Suppose there is a $(r, cr, p_1, p_2)$-sensitive family $\mathcal{H}$ for $(X, D)$, where $p_1, p_2 \in (0, 1)$ and let
    $\rho = \ln(1/p_1)/\ln(1/p_2)$.
    Then there exists a data structure for $(c, r)$-ANN over a set $P \subseteq X$ of at most $n$ points, such that:
    \begin{itemize}
        \item the query procedure requires at most $O(n^{\rho} / p_1)$ distance computations and at most $O(n^{\rho} / p_1 \cdot \lceil \log_{1 / p_2} n \rceil)$
        evaluations of the hash functions from $\mathcal{H}$ or other operations;
        \item the data structure uses at most $O(n^{1 + \rho} / p_1)$ words of space, in addition to the space needed to store the set~$P$.
    \end{itemize}
    The failure probability of the data structure can be made to be arbitrarily small constant.
\end{theorem}

\textbf{Remark:} this theorem says that in order to construct a good data structure for the $(c, r)$-ANN
it is sufficient to have a $(r, cr, p_1, p_2)$-sensitive family $\mathcal{H}$ with small
$\rho = \ln(1/p_1)/\ln(1/p_2)$ and not too small $p_1$.

We use the LSH family crafted in~\cite{ai-nohaa-06}. The properties of this family
that we need are summarized in the following theorem.

\begin{theorem}[\cite{ai-nohaa-06}]
    \label{ball_carving}
    For every sufficiently large $d$ and $n$ there exists a family $\mathcal{H}$
    of hash functions for $\ell_2^d$ such that
    \begin{itemize}
        \item a function from $\mathcal{H}$ can be sampled in time, stored in
        space, and computed in time $t^{O(t)} \cdot \log n + O(dt)$,
        where $t = \log^{2/3} n$;
        \item the collision probability of $\mathcal{H}$ for two points
        $u, v \in \Rbb^d$ depends only on the distance between $u$ and $v$;
        let us denote it by $p(\|u - v\|)$;
        \item one has the following inequalities for $p(\cdot)$:
\[
		  \begin{array}{rclrrcl}
            p(1) & \geq & L,&\hbox{where}& L & = & \frac{A}{2 \sqrt{t}} \cdot \frac{1}{(1 + \eps + 8\eps^2)^{t/2}}\\
            \forall c > 1 \quad p(c) & \leq & U(c),&\hbox{where}&
            U(c) & = & \frac{2}{(1 + c^2 \eps)^{t/2}}, 
        \end{array}
\]
        where $A$ is an absolute positive constant that is less than $1$, and
        $\eps = \Theta(t^{-1/2}) = \Theta(\log^{-1/3} n)$.
    \end{itemize}
\end{theorem}

Combining Theorem~\ref{lsh_to_nn} and Theorem~\ref{ball_carving} one has the following corollary.

\begin{corollary}
    There exists a data structure for $(c, r)$-ANN for $\ell_2^d$
    with preprocessing time and space $O_c(n^{1 + 1/c^2 + o_{c}(1)} + nd)$ and query time
    $O_c(d n^{1/c^2 + o_{c}(1)})$.
\end{corollary}
\begin{proof}
    By rescaling one can assume w.l.o.g. that $r = 1$. Then, it is left to check that
    $L = n^{-o_{c}(1)}$ and
    $
        \ln (1 / L) / \ln (1 / U(c)) \leq 1 / c^2 + o_{c}(1).
    $
    These computations can be found in~\cite{ai-nohaa-06}.
\end{proof}

We use the following standard estimate on tails of Gaussians (see, e.g.,~\cite{kms-agcsp-98}).

\begin{lemma}[\cite{kms-agcsp-98}]
    \label{gaussians}
    For every $t > 0$
    $$
        \frac{1}{\sqrt{2 \pi}} \cdot \left(\frac{1}{t} - \frac{1}{t^3}\right)
        \cdot e^{-t^2 / 2} \leq \mathrm{Pr}_{X \sim N(0, 1)}[X \geq t]
        \leq \frac{1}{\sqrt{2 \pi}} \cdot \frac{1}{t} \cdot e^{-t^2 / 2}.
    $$
\end{lemma}

We use Johnson-Lindenstrauss dimension reduction procedure.

\begin{theorem}[\cite{jl-elmhs-84}, \cite{dg-eptjl-03}]
    \label{jl_transform}
    For every $d \in \Nbb$ and $\eps, \delta > 0$ there exists a distribution over linear maps $A \colon \Rbb^d \to \Rbb^{O(\log(1 / \delta) / \eps^2)}$
    such that for every $x \in \Rbb^d$ one has
    $
        \mathrm{Pr}_A[\|Ax\| \in (1 \pm \eps) \|x\|] \geq 1 - \delta.
    $
    Moreover, such a map can be sampled in time $O(d \log(1 / \delta) / \eps^2)$.
\end{theorem}

Finally, let us state Jung's theorem.

\begin{theorem}[see, e.g., Exercise~1.3.5 in~\cite{m-ldg-02}]
    \label{jung}
    Every subset of $\Rbb^d$ with diameter $\Delta$ can be enclosed in a ball of radius $\Delta / \sqrt{2}$.
\end{theorem}

    \section{Gaussian LSH}

\label{gaussian_section}

In this section we present and analyze a $(1, c, p_1, p_2)$-sensitive family of hash functions for the $\ell_2$ norm
that gives an improvement upon~\cite{ai-nohaa-06} for the case, when
all the points and queries lie on a spherical shell of radius $O(c)$ and width $O(1)$.
The construction is similar to an SDP rounding scheme
from~\cite{kms-agcsp-98}.

First, we present an ``idealized'' family. In the following theorem we do not care about time and space complexity
and assume that all points lie on a \emph{sphere} of radius $O(c)$.

\begin{theorem}
    \label{gaussian_lsh}
    For a sufficiently large $c$, every $\nu \geq 1/2$ and $1/2 \leq \eta \leq \nu$
    there exists an LSH family for
    $\eta c \cdot S^{d-1} = \set{x \in \Rbb^d \mid \|x\| = \eta c}$ with the $\ell_2$ norm
    that is $(1, c, p_1, p_2)$-sensitive, where
    \begin{itemize}
        \item $p_1 = \exp(-o_{c,\nu}(d))$;
        \item one has
            $$
                \rho = \frac{\ln(1 / p_1)}{\ln(1 / p_2)} = \left(1 - \frac{1}{4 \eta^2}\right) \cdot \frac{1}{c^2} +
                O_{\nu}\left(\frac{1}{c^3}\right) + o_{c, \nu}(1).
            $$
    \end{itemize}
\end{theorem}
\begin{proof}
    Let $\eps > 0$ be a positive parameter that depends on $d$ as follows: $\eps = o(1)$ and $\eps = \omega(d^{-1/2})$.
    Let $\mathcal{H}$ be a family of hash functions described by Algorithm~\ref{gaussian_partitioning}
    (the pseudocode describes how to sample $h \sim \mathcal{H}$).
    \begin{algorithm}
        \caption{Gaussian partitioning}
        \label{gaussian_partitioning}
        \begin{algorithmic}
            \State $\mathcal{P} \gets \emptyset$
            \Comment{eventually, $\mathcal{P}$ will be a partition of $\eta c \cdot S^{d - 1}$}
            \While{$\bigcup \mathcal{P} \ne \eta c \cdot S^{d-1}$}
                \Comment{we denote $\bigcup \mathcal{P}$ the union of all sets that belong to $\mathcal{P}$}
                \State sample $w \sim N(0, 1)^d$
                \State $S \gets \set{u \in \eta c \cdot S^{d-1} \mid \langle u, w \rangle \geq \eta c \cdot \eps \sqrt{d}} \setminus\bigcup \mathcal{P}$
                \If{$S \ne \emptyset$}
                    \State $\mathcal{P} \gets \mathcal{P} \cup \set{S}$
                \EndIf
            \EndWhile
            \State define $h$ to be the function that maps a point $u \in \eta c \cdot S^{d-1}$ to the part of $\mathcal{P}$ that it belongs to
        \end{algorithmic}
    \end{algorithm}

    Clearly for $u, v \in \eta c \cdot S^{d-1}$ with angle $\alpha$ between them
    \begin{multline}
        \label{coll}
        \mathrm{Pr}_{h \sim \mathcal{H}}[h(u) = h(v)] = \frac{\mathrm{Pr}_{w \sim N(0, 1)^d}[\langle u, w\rangle \geq \eta c \cdot \eps \sqrt{d}
            \wedge
            \langle v, w\rangle \geq \eta c \cdot \eps \sqrt{d}]}{\mathrm{Pr}_{w \sim N(0, 1)^d}[\langle u, w\rangle \geq \eta c \cdot \eps \sqrt{d}
            \vee
            \langle v, w\rangle \geq \eta c \cdot \eps \sqrt{d}]} \\ = \Theta(1) \cdot
            \frac{\mathrm{Pr}_{X, Y \sim N(0, 1)}[X \geq \eps \sqrt{d} \wedge \cos \alpha \cdot X - \sin \alpha \cdot Y \geq \eps \sqrt{d}]}{\mathrm{Pr}_{X \sim N(0, 1)}[X \geq \eps \sqrt{d}]}\\ = \Theta(\eps \sqrt{d}) \cdot
            \frac{\mathrm{Pr}_{X, Y \sim N(0, 1)}[X \geq \eps \sqrt{d} \wedge \cos \alpha \cdot X - \sin \alpha \cdot Y \geq \eps \sqrt{d}]}{e^{-\eps^2 d / 2}}.
    \end{multline}

    In the last equality we use Lemma~\ref{gaussians} and the fact that $\eps = \omega(d^{-1/2})$.

    The following two lemmas allow us to estimate the numerator of the
    right-hand side of~(\ref{coll}).

    \begin{lemma}
        \label{numerator_up}
        $$
            \mathrm{Pr}_{X, Y \sim N(0, 1)}[X \geq \eps \sqrt{d} \wedge \cos \alpha \cdot X - \sin \alpha \cdot Y \geq \eps \sqrt{d}]
            = O\left(\frac{e^{-\eps^2 d \cdot (1 + \tan^2 \frac{\alpha}{2}) / 2}}{\eps \sqrt{d}}\right).
        $$
    \end{lemma}
    \begin{proof}
        \begin{multline*}
            \mathrm{Pr}_{X, Y \sim N(0, 1)}[X \geq \eps \sqrt{d} \wedge \cos \alpha \cdot X - \sin \alpha \cdot Y \geq \eps \sqrt{d}]
            \\ \leq \mathrm{Pr}_{X, Y \sim N(0, 1)}[(1 + \cos \alpha) \cdot X - \sin \alpha \cdot Y \geq 2 \eps \sqrt{d}]
            \\ = \mathrm{Pr}_{Z \sim N(0, 1)}[ \sqrt{(1 + \cos \alpha)^2 + \sin^2 \alpha} \cdot Z \geq 2 \eps \sqrt{d}]
            \\ = \mathrm{Pr}_{Z \sim N(0, 1)}[\sqrt{2 \cdot (1 + \cos \alpha)} \cdot Z \geq 2 \eps \sqrt{d}]
             = O\left(\frac{e^{-\eps^2 d \cdot (1 + \tan^2 \frac{\alpha}{2}) / 2}}{\eps \sqrt{d}} \right)
        \end{multline*}
        In the last equality we used Lemma~\ref{gaussians}, the fact that $\eps = \omega(d^{-1/2})$ and the identity
        $            \frac{2}{1 + \cos \alpha} = 1 + \tan^2 \frac{\alpha}{2}.
        $
    \end{proof}
    \begin{lemma}
        \label{numerator_lo}
        If $0 \leq \alpha < \alpha_0$ for some \emph{constant} $0 < \alpha_0 < \pi / 2$, then
        $$
            \mathrm{Pr}_{X, Y \sim N(0, 1)}[X \geq \eps \sqrt{d} \wedge \cos \alpha \cdot X - \sin \alpha \cdot Y \geq \eps \sqrt{d}]
            = \Omega\left(\frac{e^{-\eps^2 d \cdot (1 + \tan^2 \frac{\alpha_0}{2}) / 2}}{\eps^2 d \cdot \tan \frac{\alpha_0}{2}}\right).
        $$
    \end{lemma}
    \begin{proof}
        \begin{multline*}
            \mathrm{Pr}_{X, Y \sim N(0, 1)}[X \geq \eps \sqrt{d} \wedge \cos \alpha \cdot X - \sin \alpha \cdot Y \geq \eps \sqrt{d}]
            \\ \geq \mathrm{Pr}_{X, Y \sim N(0, 1)}[X \geq \eps \sqrt{d} \wedge Y \leq - \tan \frac{\alpha}{2} \cdot \eps \sqrt{d}] \\ =
            \mathrm{Pr}_{X \sim N(0, 1)}[X \geq \eps \sqrt{d}]
            \cdot \mathrm{Pr}_{Y \sim N(0, 1)}[Y \geq \tan \frac{\alpha}{2} \cdot
            \eps \sqrt{d}] \\ \geq
            \mathrm{Pr}_{X \sim N(0, 1)}[X \geq \eps \sqrt{d}]
            \cdot \mathrm{Pr}_{Y \sim N(0, 1)}[Y \geq \tan \frac{\alpha_0}{2} \cdot
            \eps \sqrt{d}]
            = \Omega\left(\frac{e^{-\eps^2 d \cdot (1 + \tan^2 \frac{\alpha_0}{2}) / 2}}
                    {\eps^2 d \cdot \tan \frac{\alpha_0}{2}}\right).
        \end{multline*}
        In the first inequality we use that for $\alpha < \alpha_0 < \pi / 2$
        the right-hand side event implies the left-hand side event.
        Indeed, 
        $$
            \cos \alpha \cdot X - \sin \alpha \cdot Y \geq \cos \alpha \cdot \eps \sqrt{d} + \sin \alpha \cdot \tan \frac{\alpha}{2} \cdot \eps \sqrt{d} = \eps \sqrt{d},
        $$
        since $\cos \alpha, \sin \alpha > 0$.
        In the last equality we used Lemma~\ref{gaussians}, the fact
        that $\alpha_0$ is constant and $\eps = \omega(d^{-1/2})$. 
    \end{proof}

    Thus, combining~(\ref{coll}), Lemma~\ref{numerator_up} and Lemma~\ref{numerator_lo}, we have the following estimates on the probability of collision.
    \begin{lemma}
        \label{closed_coll}
        One has
        $$
            \ln \frac{1}{\mathrm{Pr}_{h \sim \mathcal{H}}[h(u) = h(v)]}
            \geq \frac{\eps^2 d}{2} \cdot \tan^2 \frac{\alpha}{2} - O(1);
        $$
        and if $\alpha < \alpha_0$ for some constant $0 < \alpha_0 < \pi / 2$, then
        $$
            \ln \frac{1}{\mathrm{Pr}_{h \sim \mathcal{H}}[h(u) = h(v)]}
            \leq \frac{\eps^2 d}{2} \cdot \tan^2 \frac{\alpha_0}{2}
            + \ln\left(\eps \sqrt{d} \cdot \tan \frac{\alpha_0}{2}\right) + O(1).
        $$
    \end{lemma}
    Since
    $$
        \tan^2 \frac{\alpha}{2} = \frac{\|u - v\|^2 / (\eta c)^2}{4 - \|u - v\|^2 / (\eta c)^2},
    $$
    by setting $\eps = d^{-1/4}$ and invoking Lemma~\ref{closed_coll} for the angles that correspond to distances $1$ and $c$, we have
    \begin{eqnarray*}
        \ln \frac{1}{p_1} & \leq & \frac{\sqrt{d}}{2} \cdot
        \frac{1 / (\eta c)^2}{4 - 1 / (\eta c)^2} + O_{c, \nu}(\ln d), \\
        \ln \frac{1}{p_2} & \geq & \frac{\sqrt{d}}{2} \cdot \frac{1 / \eta^2}
        {4 - 1 / \eta^2} - O(1).
    \end{eqnarray*}
    Note that here we use that $c$ is large enough, since we must have $\alpha_0 < \pi / 2$ in order to be able to
    apply Lemma~\ref{closed_coll}.

    Thus, we have $p_1 = \exp(-o_{c, \nu}(d))$. A similar estimate holds for $p_2$ provided that $\eta$ is
    separated from $1/2$ (but we do not really need it). Therefore
    $$
        \rho = \frac{\ln(1 / p_1)}{\ln(1 / p_2)} = \frac{4 - 1 / \eta^2}{4 - 1 / (\eta c)^2} \cdot \frac{1}{c^2} + o_{c, \nu}(1) \\= \left(1 - \frac{1}{4 \eta^2}\right) \cdot \frac{1}{c^2} + O_{\nu}\left(\frac{1}{c^3}\right) + o_{c,\nu}(1).
    $$
\end{proof}
\textbf{Remark:} we could have had $O_{\nu}(1 / c^4)$ term in the expression for $\rho$, but we state the theorem
with $O_{\nu}(1 / c^3)$ in order to be consistent with the next theorem.

Now we show how to convert this ``idealized'' family to a real one.

\begin{theorem}
    \label{non_ideal_gaussian_lsh}
    For a sufficiently large $c$, every $\nu \geq 1/2$ and $1/2 \leq \eta \leq \nu$
    there exists an LSH family $\mathcal{H}$ for
    $$
        \set{x \in \Rbb^d \mid \|x\| \in [\eta c - 1; \eta c + 1]}
    $$
    with the $\ell_2$ norm such that
    \begin{itemize}
        \item it satisfies the conclusion of Theorem~\ref{gaussian_lsh};
        \item for every $k \in \Nbb$ one can sample a function from $\mathcal{H}$
        in time $\exp(o(d))$,
        store it in space $\exp(o(d))$ and query in time
        $\exp(o(d))$.
    \end{itemize}
\end{theorem}
\begin{proof}
    We use the family from the proof of Theorem~\ref{gaussian_lsh}, but with two
    modifications. First, if we want to compute $h(x)$ for $h \sim \mathcal{H}$,
    then before doing so, we normalize $x$ to the length $\eta c$.
    Second, in Algorithm~\ref{gaussian_partitioning} instead of checking
    the condition $\bigcup \mathcal{P} = \eta c \cdot S^{d-1}$, we simply run the partitioning
    process for $\exp(o(d))$ steps. Namely, we require that
    after the end the probability of the event $\bigcup \mathcal{P} = \eta c \cdot S^{d-1}$
    is at least $1 - \exp(-d)$ (one can see that this will be the case
    after $\exp(o(d))$ steps by a standard $\eps$-net argument).
    Such a high probability means that this LSH family achieves the same parameters as
    the one from Theorem~\ref{gaussian_lsh}.
    Clearly, such a function can be stored in space $\exp(o(d))$
    and queried in time $\exp(o(d))$.

    It is left to argue that normalizing a vector before computing $h$
    does not affect the quality (namely, we are interested in $p_1$
    and $\rho$) by a lot.

\begin{lemma}
For any vectors $u$ and $v$,
$$\|u/\|u\| - v/\|v\|\|^2 = \frac{1}{\|u\|\cdot\|v\|}\left(\|u-v\|^2 - (\|u\|-\|v\|)^2\right)$$
\end{lemma}
\begin{proof}
  \begin{align*}    
    \|u/\|u\| - v/\|v\|\|^2 &= 2 - \frac{2\langle u,v\rangle}{\|u\|\cdot\|v\|}
    =\frac{1}{\|u\|\cdot\|v\|}\left(\|u-v\|^2 - (\|u\|-\|v\|)^2\right).
  \end{align*}
\end{proof}

    By the above lemma, one can check that for $u, v \in \set{x \in \Rbb^d \mid \|x\| \in [\eta c - 1; \eta c + 1]}$
    \begin{itemize}
        \item if $\|u - v\| \leq 1$, then
$
            (\eta c \cdot \|u / \|u\| - v / \|v\|\|)^2 \le \frac{(\eta c)^2}{(\eta c - 1)^2}
            \le 1 + O_{\nu}\left(\frac1c\right)
$
        \item if $\|u - v\| \geq c$, then
$ (\eta c \cdot \|u / \|u\| - v / \|v\|\|)^2 \ge \frac{(\eta c)^2}{(\eta c + 1)^2}(c^2 - 4)
            \ge c^2 \cdot \left(1 - O_{\nu}\left(\frac1c\right)\right).
$
    \end{itemize}
    Clearly, from these inequalities we can see that the conclusion of
    Theorem~\ref{gaussian_lsh} is still true for our case.
\end{proof}

    \section{Two-level hashing}

\begin{algorithm}
    \caption{Two-level hashing}
    \label{two_level_hashing_alg}
    \begin{algorithmic}[1]
        \Function{Build}{$P$, $\tau$, $T$, $k$, $\widetilde{k}_l$}
            \State sample $h \sim \mathcal{H}_1^{\otimes k}$, where $\mathcal{H}_1$ is a family from Theorem~\ref{ball_carving} (w.l.o.g. $h$ maps $\Rbb^d$ to $[m]$)
            \State $B_i \gets \set{p \in P \mid h(p) = i}$
            \For{$i \gets 1 \ldots m$}
                \While{there exists $p_1, p_2 \in B_i$ such that $\|p_1 - p_2\| > \tau c$}
                    \State $B_i \gets B_i \setminus \set{p_1, p_2}$
                \EndWhile
                \If{$B_i \ne \emptyset$}
                    \State let $u_i$ be the center of the smallest enclosing ball of $B_i$
                    \State let $s_i \in B_i$ be the nearest neighbor of $u_i$
                    \For{$l \gets 0 \ldots T$}
                        \State $\widetilde{P}_{il} \gets \set{p - u_i \mid p \in B_i, c/2 + l - 1 \leq \|p - u_i\| \leq c/2 + l + 1}$
                        \State sample $\widetilde{h}_{il} \sim \mathcal{H}_2^{\otimes \widetilde{k}_l}$, where $\mathcal{H}_2$ is a family from Theorem~\ref{non_ideal_gaussian_lsh}
                        for $\eta = 1/2 + l/c$
                        \State $\widetilde{B}_{ilj} \gets \set{p \in \widetilde{P}_{il} \mid \widetilde{h}_{il}(p) = j}$
                    \EndFor
                \EndIf
            \EndFor
        \EndFunction
        \Function{Query}{$q$, $T$}
            \State $i \gets h(q)$
            \If{$B_i = \emptyset$}
                \State\Return{$\perp$}
            \EndIf
            \If{$\|q - s_i\| \leq c$}
                \State \Return{$s_i$}
            \EndIf
            \For{$l \gets 0 \ldots T$}
                \If{$c/2 + l - 1 \leq \|q - u_i\| \leq c/2 + l + 1$}
                    \State $j \gets \widetilde{h}_{il}(q - u_i)$
                    \For{$p \in \widetilde{B}_{ilj}$}
                        \If{$\|q - (p + u_i)\| \leq c$}
                            \State \Return{$p + u_i$}
                        \EndIf
                    \EndFor
                \EndIf
            \EndFor
            \State \Return{$\perp$}
        \EndFunction
    \end{algorithmic}
\end{algorithm}

We now describe our near neighbor data structure. It is composed of
several independent data structures, where each one is a two-level
hashing scheme, described next. We will conclude with proving our main
theorem for ANN search.

First, we provide some very high-level intuition.

\subsection*{Intuition}

The general approach can be seen as using LSH scheme composed of two
levels: the hash function is $h=(h_C, h_G)$ chosen from two families
$h_C\in {\cal H}^{k'}, h_G\in {\cal G}^l$ for some $k',l$, where $\cal
H$ is the ``ball carving LSH'' from Theorem \ref{ball_carving}, and
$\cal G$ is the ``Gaussian LSH'' from Theorem
\ref{non_ideal_gaussian_lsh}. In particular, the hash function
$h_G(p)$ will depend on the bucket $h_C(p)$ and the other dataset
points in the bucket $h_C(p)$. Intuitively, the ``outer level'' hash
function $h_C$ performs a partial standard LSH partitioning (with $\rho\approx
1/c^2$), but also has the role of improving the ``geometry'' of the
points (namely, the points in a buckets roughly will have a bounded
diameter). After an application of $h_C$, the
pointset (inside a fixed bucket) has bounded diameter, allowing us to
use the improved Gaussian LSH partitioning (``inner level''), with
$\rho<1/c^2$.

In more detail, first, let us recall the main idea of the proof of
Theorem~\ref{lsh_to_nn}.  Suppose that $\mathcal{H}$ is, say, a family
from Theorem~\ref{ball_carving}.  Then, we choose $k$ to be an integer
such that for every $p, q \in \Rbb^d$ with $\|p - q\| \geq c$ we
have
\begin{equation}
    \label{far_prob}
    \mathrm{Pr}_{h \sim \mathcal{H}^{\otimes k}}[h(p) = h(q)] \approx n^{-1}.
\end{equation}
Then, by Theorem~\ref{ball_carving}, we have for every $p, q \in \Rbb^d$ with
$\|p - q\| \leq 1$
$$
    \mathrm{Pr}_{h \sim \mathcal{H}^{\otimes k}}[h(p) = h(q)] \approx n^{-1/c^2}.
$$ 

Now suppose we hash all the points using a function $h \sim
\mathcal{H}^{\otimes k}$.  For a query $q$ the average number of
``outliers'' in a bin that corresponds to $q$ (points $p$ such that
$\|p - q\| > c$) is at most $1$ due to~(\ref{far_prob}).  On the other
hand, for a data point $p$ such that $\|p - q\| \leq 1$ the
probability of collision is at least $n^{-1 / c^2}$.  Thus, we can
create $n^{1/c^2}$ independent hash tables, and query them all in time
around $O(n^{1/c^2})$.  The resulting probability of success is
constant.

The above analysis relies on two distance scales: $1$ and $c$. To get
a better algorithm for ANN we introduce the third scale: $\tau c$ for
$\tau > 1$ being a parameter.  First, we hash all the points using
$\mathcal{H}^{\otimes k'}$ (where $k' \approx k / \tau$) so that the collision probabilities are
roughly as follows.

\begin{center}
\begin{tabular}{|r|c|c|c|}
    \hline
    Distance & $1$ & $c$ & $\tau c$\\
    \hline
    Probability of collision & $n^{-1/(\tau c)^2}$ & $n^{-1 / \tau^2}$ & $n^{-1}$\\
    \hline
\end{tabular}
\end{center}

Now we can argue that with high probability any bucket has diameter $O_{\tau}(c)$.
This allows us to use the family from Theorem~\ref{non_ideal_gaussian_lsh} for each bucket and set
probabilities of collision as follows.

\begin{center}
\begin{tabular}{|r|c|c|}
    \hline
    Distance & $1$ & $c$ \\
    \hline
    Probability of collision & $n^{-(1 - \Omega_{\tau}(1)) \cdot (1 - 1 / \tau^2) / c^2}$ & $n^{-1 + 1 / \tau^2}$\\
    \hline
\end{tabular}
\end{center}

Due to the independence, we expect overall collision probabilities to be products
of ``outer'' collision probabilities from the first table and ``inner'' probabilities
from the second table.
Thus, in total, we have the following probabilities.
\begin{center}
\begin{tabular}{|r|c|c|}
    \hline
    Distance & $1$ & $c$ \\
    \hline
    Probability of collision & $n^{-(1 - \Omega_{\tau}(1))/ c^2}$ & $n^{-1}$\\
    \hline
\end{tabular}
\end{center}
Then we argue as before and achieve
$$
    \rho \approx \frac{1 - \Omega_{\tau}(1)}{c^2}.
$$

This plan is not quite rigorous for several reasons.
One of them is we do not properly take care of conditioning on the event ``all buckets have low diameter''.
Nevertheless, in this section we show how to analyze a similar scheme rigorously.

\subsection*{Construction}

We want to solve $(c, 1)$-ANN for $\ell_2^d$.
As a first step, we apply Johnson-Lindenstrauss transform (Theorem~\ref{jl_transform})
and reduce our problem to $(c - 1, 1)$-ANN for $\ell_2^{O_{c}(\log n)}$
by increasing the failure probability by an arbitrarily small constant.
This means that all quantities of order $\exp(o(d))$ are now $n^{o_c(1)}$
(in particular, various parameters of the hash family from Theorem~\ref{non_ideal_gaussian_lsh}).
Abusing notation, let us assume that we are solving $(c, 1)$-ANN in $\ell_2^{O_c(\log n)}$.

For the description of preprocessing and query algorithms see Algorithm~\ref{two_level_hashing_alg}.
Roughly speaking, we first hash points using a hash family from Theorem~\ref{ball_carving}
and then for every bucket we utilize a family from Theorem~\ref{non_ideal_gaussian_lsh} (after some pruning).
The hashing scheme has several parameters: $\tau$, $T$, $k$ and $\widetilde{k}_l$ for $0 \leq l \leq T$. Let us show how to set them.
First, we choose some $\tau > 1$ (we will set a concrete value later).
Second, we choose
\begin{equation}
    \label{choice_t}
    T = \left\lceil \frac{\tau c}{\sqrt{2}} - \frac{c}{2} \right\rceil + 1.
\end{equation}
Third, we choose $k$ to be smallest positive integer such that
\begin{equation}
    \label{choice_k}
    \left(\frac{U(\tau c - 1)}{L}\right)^k \leq \frac{1}{2n},
\end{equation}
where $U(\cdot)$ and $L$ are from Theorem~\ref{ball_carving}.
Finally, for every $0 \leq l \leq T$ we set $\widetilde{k}_l$ to be the smallest positive integer such that
for every $u, v \in \Rbb^d$ with $\|p_1\|, \|p_2\| \in \left[c / 2 + l - 1; c/2 + l + 1\right]$ and $\|p_1 - p_2\| \geq c$
we have
\begin{equation}
    \label{choice_kl}
    U(c)^{k} \cdot \mathrm{Pr}_{\widetilde{h} \sim \mathcal{H}_2^{\otimes \widetilde{k}_l}}\left[\widetilde{h}(u) = \widetilde{h}(v)\right] \leq \frac{1}{3n},
\end{equation}
where $\mathcal{H}_2$ is a family from Theorem~\ref{non_ideal_gaussian_lsh} for $\eta = 1/2 + l/c$.

It is immediate to see that, if the query algorithm outputs some point $p$, then $p \in P$ and $p$ is within distance $c$ from a query.

\subsection*{Auxiliary lemmas}

\begin{lemma}
    \label{form_partition}
    After the preprocessing one has for every $1 \leq i \leq m$
    $$
        B_i = \bigcup_{0 \leq l \leq T} \widetilde{P}_{il}.
    $$
\end{lemma}
\begin{proof}
    This follows from Jung's theorem (Theorem~\ref{jung}). 

    Indeed, after the lines 5--7 the diameter of $B_i$ is at most $\tau c$.
    Thus, the radius of the smallest enclosing ball is at most $\tau c / \sqrt{2}$.
    It means that for any $l > T$ the set $\widetilde{P}_{il}$ is empty.
\end{proof}

\begin{lemma}
    \label{prob_ineq}
    If $\mathcal{U}$ and $\mathcal{V}$ are two events
    with $\mathrm{Pr}[\mathcal{V}] < 1$, then
    $$
        \mathrm{Pr}[\mathcal{U} \vee \mathcal{V}] \geq \mathrm{Pr}[\mathcal{U} \mid \neg \mathcal{V}].
    $$
\end{lemma}
\begin{proof}
    \begin{align*}
        \Prb{}{\Uc \vee \Vc} &= \Prb{}{\Vc} + \Prb{}{\Uc \mid \neg \Vc} \Prb{}{\neg \Vc}\\
        &\ge  \Prb{}{\Uc \mid \neg \Vc}  \Prb{}{\Vc} + \Prb{}{\Uc \mid \neg \Vc} \Prb{}{\neg \Vc}\\
        &=\Prb{}{\Uc \mid \neg \Vc}
    \end{align*}
\end{proof}

\subsection*{Collision probabilities}

Suppose that $q \in \Rbb^d$ is a query and $p \in P$ is a data point.
Let us introduce four events:
\begin{itemize}
    \item $\mathcal{A}$ stands for ``$h(p) = h(q)$'';
    \item $\mathcal{B}$: ``for every $p' \in P$ such that $\|p' - q\| > \tau c - 1$ we have $h(p') \ne h(q)$'';
    \item $\mathcal{C}$: ``we iterate through $p$ in the line 30 of Algorithm~\ref{two_level_hashing_alg}'';
    \item $\mathcal{D}$: ``$B_{h(q)} \ne \emptyset$ and $\|q - s_{h(q)}\| \leq c$''.
\end{itemize}

\begin{lemma}
    \label{upper_coll_lemma}
    If $\|p - q\| \geq c$, then
    \begin{equation}
        \label{upper_coll}
        \mathrm{Pr}[\mathcal{C}] \leq 1 / n.
    \end{equation}
\end{lemma}
\begin{proof}
    Since $\Cc$ implies $\Ac$, we have
    \begin{equation}
        \label{short_chain}
        \Prb{}{\Cc} = \Prb{}{\Ac} \Prb{}{\Cc \mid \Ac}.
    \end{equation}
    By Theorem~\ref{ball_carving} $\Prb{}{\Ac} \leq U(c)^k$.
    Moreover, if we denote $$W_q = \set{l \in \Zbb_+ \mid c / 2 + l - 1 \leq \|q - u_{h(q)} \| \leq c / 2 + l + 1},$$ then
    $$
        \Prb{}{\Cc \mid \Ac} = \sum_{l \in W_q} \Prb{}{p \in \widetilde{P}_{h(q) l} \wedge \widetilde{h}_{h(q) l}(p) = \widetilde{h}_{h(q) l}(q)}.
    $$
    Since $|W_q| \leq 3$ and due to~(\ref{short_chain}) and~(\ref{choice_kl}) we have~(\ref{upper_coll}).
\end{proof}

\begin{lemma}
    \label{lower_coll_lemma}
    If $\|p - q\| \leq 1$, then
    $$
        \mathrm{Pr}[\mathcal{C} \vee \mathcal{D}] \geq
        n^{-\left(1 - \frac{1}{2 \tau^2} + \frac{1}{2 \tau^4}\right) \cdot \frac{1}{c^2} + O_{\tau}\left(\frac{1}{c^3}\right) + o_{c, \tau}(1)}.
    $$
\end{lemma}
\begin{proof}
    Using Lemma~\ref{prob_ineq}, we get
    \begin{multline}
        \label{long_chain}
        \Prb{}{\Cc \vee \Dc} \geq \Prb{}{\Ac \wedge \Bc \wedge (\Cc \vee \Dc)} = \Prb{}{\Ac} \Prb{}{\Bc \mid \Ac} \Prb{}{\Cc \vee \Dc \mid \Ac \wedge \Bc} 
        \\ \geq \Prb{}{\Ac} \Prb{}{\Bc \mid \Ac} \Prb{}{\Cc \mid \Ac \wedge \Bc \wedge \neg \Dc}.
    \end{multline}
    In the following three lemmas we lower bound the right-hand side of~(\ref{long_chain}).
    \begin{lemma}
        \label{prob_a}
        $\Prb{}{\Ac} \geq L^{k}$
    \end{lemma}
    \begin{proof}
        This follows immediately from Theorem~\ref{ball_carving}.
    \end{proof}
    \begin{lemma}
        \label{prob_b}
        $\Prb{}{\Bc \mid \Ac} \geq 1/2$
    \end{lemma}
    \begin{proof}
        \begin{multline*}
            \Prb{}{\neg \Bc \mid \Ac} \leq \sum_{\begin{smallmatrix}p' \in P \\ \|p' - q\| > \tau c - 1\end{smallmatrix}} \Prb{}{h(p') = h(q) \mid h(p) = h(q)} \\ \leq \sum_{\begin{smallmatrix}p' \in P \\ \|p' - q\| > \tau c - 1\end{smallmatrix}}
            \frac{\Prb{}{h(p') = h(q)}}{\Prb{}{h(p) = h(q)}} \leq n \cdot \left(\frac{U(\tau c - 1)}{L}\right)^k \leq
            1 / 2,
        \end{multline*}
        where the penultimate inequality is due to Lemma~\ref{prob_a}, and the last one is due to~(\ref{choice_k}).
    \end{proof}
    \begin{lemma}
        \label{prob_c}
        $$
            \Prb{}{\Cc \mid \Ac \wedge \Bc \wedge \neg \Dc} \geq \left(\frac{1}{3n \cdot U(c)^k}\right)^{\left(1 - \frac{1}{2\tau^2}\right)\cdot \frac{1}{c^2} + O_{\tau}\left(\frac{1}{c^3}\right) + o_{c, \tau}(1)}
        $$
    \end{lemma}
    \begin{proof}
        First, since we condition on $\Ac \wedge \Bc$, we have that $p \in B_{h(q)}$ after the preprocessing.
        Indeed, $\Ac$ implies that $p \in B_{h(q)}$ in the beginning of the preprocessing,
        and $\Bc$ together with $\|p - q\| \leq 1$ imply that this will be the case in the end as well.

        Second, let us prove that $\|p - u_{h(q)}\| \geq c / 2 - 1$ and $\|q - u_{h(q)}\| \geq c/2 - 1$.
        Indeed, 
        \begin{multline*}
            c \leq \|q - s_{h(q)}\| \leq \|q - u_{h(q)}\| + \|s_{h(q)} - u_{h(q)}\|
            \leq \|q - u_{h(q)}\| + \|p - u_{h(q)}\| \\ \leq
            2 \cdot \|p - u_{h(q)}\| + \|p - q\| \leq 2 \cdot \|p - u_{h(q)}\| + 1,
        \end{multline*}
        where the first inequality follows from $\neg \Dc$, and the third inequality follows from the fact that $p$ is not filtered from $B_{h(q)}$ and the definition
        of $s_{h(q)}$.
        As a result, we have $\|p - u_{h(q)}\| \geq (c - 1) / 2 \geq c / 2 - 1$. Similarly, we prove
        $\|q - u_{h(q)}\| \geq c / 2 - 1$.

        As a result, we have that for some $0 \leq l \leq T$, $p \in \widetilde{P}_{h(q) l}$ and, moreover,
        $$
            c/2 + l - 1 \leq \|q - u_{h(q)}\| \leq c/2 + l + 1,
        $$
        so 
        $$
            \Prb{}{\Cc \mid \Ac \wedge \Bc \wedge \neg \Dc} \geq \Prb{}{\widetilde{h}_{h(q)l}(p) = \widetilde{h}_{h(q)l}(q)},
        $$
        but due to Theorem~\ref{non_ideal_gaussian_lsh}, (\ref{choice_kl}), (\ref{choice_t}) and the minimality of $k_l$ this implies
        \begin{equation}
            \label{main_claim}
            \Prb{}{\Cc \mid \Ac \wedge \Bc \wedge \neg \Dc} \geq \left(\frac{1}{3n \cdot U(c)^k}\right)^{\left(1 - \frac{1}{2\tau^2}\right)\cdot \frac{1}{c^2} + O_{\tau}\left(\frac{1}{c^3}\right) + o_{c, \tau}(1)},
        \end{equation}
        since we can apply Theorem~\ref{non_ideal_gaussian_lsh} for
        $$
            \eta = \frac{1}{2} + \frac{l}{c} \leq \frac{1}{2} + \frac{T}{c}
            \leq \frac{\tau}{\sqrt{2}} + O_{\tau}\left(\frac{1}{c}\right),
        $$
        and as a result we have
        $$
            \rho \leq \left(1 - \frac{1}{4 \eta^2}\right) \cdot \frac{1}{c^2} + O_{\tau}\left(\frac{1}{c^3}\right) + o_{c, \tau}(1)
            \leq \left(1 - \frac{1}{2 \tau^2}\right) \cdot \frac{1}{c^2} + O_{\tau}\left(\frac{1}{c^3}\right) + o_{c,\tau}(1),
        $$
        which in turn implies~(\ref{main_claim}).
    \end{proof}

    Now we can finish the proof of Lemma~\ref{lower_coll_lemma}.
    Combining Lemmas~\ref{prob_a}, \ref{prob_b} and \ref{prob_c} we get
    \begin{equation}
        \label{cd_intermediate}
        \Prb{}{\Cc \vee \Dc} \geq
        L^k \cdot 
            \left(\frac{1}{n \cdot U(c)^k}\right)^{\left(1 - \frac{1}{2\tau^2}\right)\cdot \frac{1}{c^2} + O_{\tau}\left(\frac{1}{c^3}\right) + o_{c, \tau}(1)}.
    \end{equation}
    It is known from~\cite{ai-nohaa-06} that
    $$
        U(x) = L^{x^2 + o_x(1)},
    $$
    so from~(\ref{choice_k}) and the minimality of $k$ we get
    \begin{equation}
        \label{l_estimate}
        L^k = n^{-\frac{1}{(\tau c - 1)^2 - 1} + o_{c, \tau}(1)} = n^{-\frac{1}{\tau^2 c^2} + O_{\tau}\left(\frac{1}{c^3}\right) + o_{c,\tau}(1)},
    \end{equation}
    and
    \begin{equation}
        \label{u_estimate}
        U(c)^k = n^{-\frac{c^2}{(\tau c - 1)^2 - 1} + o_{c, \tau}(1)} = n^{-\frac{1}{\tau^2} + O_{\tau}\left(\frac{1}{c}\right) + o_{c,\tau}(1)}.
    \end{equation}
    Combining~(\ref{cd_intermediate}), (\ref{l_estimate}) and~(\ref{u_estimate}) we get
    $$
        \Prb{}{\Cc \vee \Dc} \geq n^{-\left(1 - \frac{1}{2 \tau^2} + \frac{1}{2 \tau^4}\right) \cdot \frac{1}{c^2}
        + O_{\tau}\left(\frac{1}{c^3}\right) + o_{c, \tau}(1)}.
    $$
\end{proof}

\subsection*{The main result}

Finally, we formulate and prove the main result.

\begin{theorem}
    \label{the_main_result}
    There exists a data structure for $(c, 1)$-ANN for $\ell_2^d$ with
    \begin{itemize}
        \item preprocessing time $O_c(n^{2 + \rho} + nd \log n)$,
        \item space $O_c(n^{1 + \rho} + d \log n)$,
        \item query time $O_c(n^{\rho} + d \log n)$,
    \end{itemize}
    where
    \begin{equation}
        \label{rho_bound}
        \rho \leq \frac{7/8}{c^2} + O\left(\frac{1}{c^3}\right) + o_c(1).
    \end{equation}
\end{theorem}
\begin{proof}
    First, we analyze the running time of one two-level hash table.

    The dimension reduction step takes time $O_c(nd \log n)$ by Theorem~\ref{jl_transform}.

    The preprocessing takes $O_c(n^{1 + o_{c, \tau}(1)})$ time plus the time needed for the lines 5--7 and the line 9.
    It is straightforward to implement lines 5--7 in time $O_c(n^2 \log n)$ and one can use
    an algorithm from~\cite{gls-gaco-88} for finding $u_i$ with running time $O_c(n \log^{O(1)} n)$.

    The query algorithm takes time $O_c(n^{o_{c, \tau}(1)} + d \log n)$ in expectation due to Lemma~\ref{upper_coll_lemma}.
    This is because in line 30 we iterate over at most $1$ point on average that is not an approximate near neighbor.

    By Lemma~\ref{lower_coll_lemma} the probability of finding an approximate near neighbor is at least
    $$
        Q =  
        n^{-\left(1 - \frac{1}{2 \tau^2} + \frac{1}{2 \tau^4}\right) \cdot \frac{1}{c^2} + O_{\tau}\left(\frac{1}{c^3}\right) + o_{\tau, c}(1)}.
    $$

    In order to make the probability of success constant we build and query $1 / Q$ independent copies of the two-level data structure.
    As a result we get the desired bounds with
    $$
        \rho = \rho(\tau) \leq \left(1 - \frac{1}{2 \tau^2} + \frac{1}{2 \tau^4}\right) \cdot \frac{1}{c^2} + O_{\tau}\left(\frac{1}{c^3}\right) + o_{\tau, c}(1).
    $$
    The bound stated in~(\ref{rho_bound}) is obtained by setting $\tau = \sqrt{2}$.

    Note that the above bound on the query time is \emph{in expectation}, but it is also possible to modify the algorithm slightly to get a worst-case bound. The algorithm still iterates over $1/Q$ tables but stops after looking at $3/Q+1$ points without finding an approximate near neighbor. The expected number of points the algorithm has to look at that are not an approximate near neighbor is at most $1/Q$. By Markov's inequality, with probability at least $2/3$, the algorithm doesn't look at more than $3/Q$ points that are not an approximate near neighbor. In each two-level table, the probability that the algorithm fails to find an approximate near neighbor is at most $1-Q$. Thus, the probability it fails in all $Q$ tables is at most $(1-Q)^{1/Q} \le 1/e$. Overall, with probability at least $1-1/3-1/e$, the algorithm finds an approximate near neighbor without looking at more than $3/Q+1$ points. 
\end{proof}

\textbf{Remark:} if one is willing to have quasi-linear preprocessing time, then it is possible to modify 
Algorithm~\ref{two_level_hashing_alg} slightly to achieve
$$
    \rho \leq \frac{15/16}{c^2} + O\left(\frac{1}{c^3}\right) + o_c(1),
$$
while having preprocessing time $O_c(n^{1 + \rho} + nd \log n)$.
The idea is to choose an arbitrary point $p \in B_i$ after the initial hashing and then remove from $B_i$ points
that are further from $p$ than $\tau c$.
After this filtering the algorithm is the same as before.
We save in preprocessing time, since we no longer need to run lines 5--7.

The details are almost the same as in the proof of Theorem~\ref{the_main_result} and thus omitted.

    \section{Acknowledgments}
H.L.N.'s work was supported by NSF CCF 0832797, an IBM Ph.D. Fellowship, and a Siebel Scholarship.
I.R. would like to thank Art\={u}rs Ba\v{c}kurs and Sepideh Mahabadi for reading the manuscript thoroughly and providing useful feedback and Akamai Presidential Fellowship for the support.

    \bibliographystyle{alpha}
    \bibliography{../../bibtex/desk}
\end{document}